\documentclass[aps,pre,twocolumn,amsmath,amssymb,superscriptaddress]{revtex4-1}
\usepackage{graphicx,color,array}
\usepackage{subfigure}
\usepackage{graphicx}
\usepackage{amsmath}
\usepackage{amsfonts}
\usepackage{amssymb}
\usepackage{latexsym,array,delarray,amsthm,epsfig}
\usepackage{multirow}
\usepackage{fancyhdr}
\usepackage{cancel}
\usepackage{paralist}
\usepackage{color}
\usepackage{hyperref}

\newcommand{\ddt}[1]{\frac{\mathrm{d} #1}{\mathrm{d} t}}
\newcommand{\pdt}[1]{\frac{\partial #1}{\partial t}}
\newcommand{\pdz}[1]{\frac{\partial #1}{\partial z}}
\newcommand{\pdr}[1]{\frac{\partial #1}{\partial r}}

\newcommand{\pdd}[2]{\frac{\partial #1}{\partial #2}}
\newcommand{\did}[2]{\frac{\mathrm{d}^{(i)}#1}{\mathrm{d}#2}}

\newcommand{\boldt}[1]{\textbf{#1}}
\newcommand{\bolds}[1]{\boldsymbol{#1}}
\newcommand{\del}{\boldsymbol{\nabla}}
\newcommand{\ud}{\mathrm{d}}
\newcommand{\uc}{\mathrm{c}}
\newcommand{\uj}{\mathrm{j}}
\newcommand{\upi}{^{(i)}}
\newcommand{\upe}{^{(e)}}
\newcommand{\nhat}{\hat{\textbf{n}}}
\newcommand{\that}{\hat{\textbf{t}}}
\newcommand{\eps}{\epsilon}
\newcommand{\rey}{\mathrm{Re}}

\begin{document}


\title{Dispersive Hydrodynamics in Viscous Fluid Conduits}

\date{\today}

\author{N.K. Lowman}
 \email{nklowman@ncsu.edu}
 \affiliation{Department of Mathematics, North Carolina State
 University, Raleigh, North Carolina 27695, USA}
\author{M. A. Hoefer}
 \affiliation{Department of Mathematics, North Carolina State
 University, Raleigh, North Carolina 27695, USA}


\begin{abstract}

The evolution of the interface separating a conduit of light, viscous fluid rising buoyantly through a heavy, more viscous, exterior fluid at small Reynolds numbers is governed by the interplay between nonlinearity and dispersion.  Previous authors have proposed an approximate model equation based on physical arguments, but a precise theoretical treatment for this two fluid system with a free boundary is lacking.  Here, a derivation of the interfacial equation via a multiple scales, perturbation technique is presented.  Perturbations about a state of vertically uniform, laminar conduit flow are considered in the context of the Navier-Stokes equations with appropriate boundary conditions.  The ratio of interior to exterior viscosities is the small parameter used in the asymptotic analysis, which leads systematically to a maximal balance between buoyancy driven, nonlinear self-steepening and viscous, interfacial stress induced, nonlinear dispersion.  This results in a scalar, nonlinear partial differential equation describing large amplitude dynamics of the cross-sectional area of the intrusive fluid conduit, in agreement with previous derivations.  The leading order behavior of the two fluid system is completely characterized in terms of the interfacial dynamics.  The regime of model validity is characterized and shown to agree with previous experimental studies.  Viscous fluid conduits provide a robust setting for the study of nonlinear, dispersive wave phenomena.

\end{abstract}

\maketitle


\section{Introduction}

Upon introduction of a steady source of light, viscous fluid into the base of a quiescent basin of dense, more viscous fluid, a diapir will form.  Once the radius of the diapir exceeds a critical threshold, the diapir will rise buoyantly through the exterior fluid, trailed by a vertically uniform, axisymmetric conduit, if its Reynolds number ($\rey$) is sufficiently low \citep{whitehead_dynamics_1975}.  Unsteady perturbations of the injection rate have been shown in a laboratory setting to produce hallmark features of nonlinear, dispersive systems including solitary waves and nonlinear wavetrains (cf. the review in \citep{ryan_magma_1990}).  To fully describe the behavior of this miscible, two-fluid, interfacial flow, one must consider the full system of governing equations with boundary conditions along a moving, free interface, a difficult task even for numerical simulations.  However, an approximate model equation has been proposed on the basis of physical arguments.  In the limit of gently sloping conduit walls, disturbances propagating upward along the interface of the conduit induced by unsteady injection are balanced by viscous forces from the exterior, resulting in wave propagation \citep{olson_solitary_1986,scott_observations_1986,helfrich_solitary_1990}. Utilizing the conduit geometry, the leading order evolution of the system can be described in terms of the (dimensionless) cross-sectional area of the conduit, $A$, by the nonlinear, dispersive, scalar partial differential equation (PDE)
\begin{equation}
	\label{eq:magma}
	\pdt{A} + \pdz{ \ }  \left\{ A^2 \left[ 1 - \pdz{ \ } \left( A^{-1} \pdt{A} \right) \right] \right\} = 0 \ , 
\end{equation}
we term the conduit equation.
The relative ease of realizing this viscous conduit setting in experiment and the analytical tractability of the model equation \eqref{eq:magma} make this an ideal platform for the study of nonlinear, dispersive waves.  

Interest in viscous fluid conduits began in the geophysics community nearly three decades ago and has continued to the present day due to the prominence of fundamental nonlinear wave phenomena.  This simplified system is thought to capture the essential physics of magma rising buoyantly along thermal plumes in the convecting mantle \cite{olson_solitary_1986,scott_observations_1986} and is closely related to interpenetrating magma flow in a viscously deformable, porous matrix where the cross-sectional area of the conduit is comparable to the matrix porosity \citep{mckenzie_generation_1984,scott_magma_1984,scott_magma_1986,simpson_multiscale_2010}.   Like in the conduit setting, the primitive equations for interpenetrating magma flow can be reduced to eq.~\eqref{eq:magma} in the proper physical setting, thus it is commonly referred to as the magma equation.  While counterintuitive at first, that in both settings, viscosity-dominated dynamics lead to a conservative (dissipationless) equation, this is natural in the conduit setting because \eqref{eq:magma} is simply an expression for mass conservation of the conduit fluid.  

In the context of viscous fluid conduits, time evolution of the conduit area is driven by a nonlinear self-steepening term due to buoyancy and a dispersive term due to viscous stress imparted by slow deformation of the exterior fluid.  Solitary waves supported by the conduit equation have been studied in detail analytically and shown to be asymptotically stable (e.g. \citep{scott_magma_1984,nakayama_rarefactive_1992,simpson_asymptotic_2008}).  These solitary waves are readily observable in experiment by the generation of a localized pulse in the rate of injection, and careful comparisons between the theoretical amplitude-speed relation of eq.~\eqref{eq:magma} and experimental data yield good agreement for small to moderate amplitude solitary waves \citep{olson_solitary_1986,scott_observations_1986,helfrich_solitary_1990}, though a precise explanation of the deviation between theory and experiment has not been identified.  The conduit equation has also been shown theoretically and numerically to produce slowly modulated, dynamically expanding, periodic wavetrains in response to nonlinear steepening, whose speeds and amplitudes can be analytically predicted using a nonlinear wave averaging technique \citep{spiegelman_flow_1993,spiegelman_flow_1993-1,lowman_dispersive_2013}.  These wavetrains correspond to dispersively regularized shock waves (DSWs)--analogous to classical, viscous shocks--which have garnered significant attention in the past decade due to their realization in optics and superfluids \citep{hoefer_dispersive_2009}.  DSWs in viscous fluid conduits have been observed experimentally \citep{scott_observations_1986,ryan_magma_1990}, but their properties have never been studied.  

A major hindrance to a more robust, quantitative study of nonlinear, dispersive waves in this system is the lack of a systematic derivation of the conduit equation from the full Navier-Stokes equations.  In this work, we present a derivation of eq.~\eqref{eq:magma} utilizing a multiple scales, perturbation approach, providing confirmation of existing intuitive arguments.  This derivation results in a complete characterization of the leading order behavior of the system (intrusive/exterior fluid velocities and pressures) in terms of $A(z,t)$.  It is found that the vertical flow in the conduit imposes a shear stress on the interface, driving a weak vertical velocity in the exterior fluid which persists far from the conduit walls, presently neglected in existing derivations.  Scaling relations between the fluid quantities are made explicit.  Model validity is characterized in terms of two independent parameters, the ratio of internal to external fluid viscosities 
\begin{equation}
  \label{eq:2}
  \eps = \mu^{(i)}/\mu^{(e)}\ ,
\end{equation}
and the internal fluid's Reynolds number $\rey^{(i)}$.  When $\eps \ll 1$ and $\rey^{(i)} = \mathcal{O}(1)$, eq.~\eqref{eq:magma} is valid for times $t \ll 1/\eps$ and amplitudes $A \ll 1/\eps$.  This translates to dimensional time units $T/\eps^{3/2}$, where the characteristic time scale $T$ is typically on the order of seconds or tenths of seconds.  In contrast to well-known models of \emph{weakly} nonlinear, interfacial fluid dynamics such as the Korteweg-deVries \citep{korteweg_change_1895} and Benjamin-Ono equations \citep{benjamin_internal_1967,ono_algebraic_1975}, the conduit equation is valid for \emph{arbitrarily large} amplitudes assuming $\eps$ is sufficiently small.
Finally, we determine the regime of model validity when inertial, external boundary, surface tension, higher order viscous, and symmetry-breaking effects are present.  

The remainder of the article will be organized in the following manner.  Section \ref{sec:problem} describes the governing equations and their nondimensionalization, as well as the normalization of the fluid properties to the appropriate scales of interest.  In \S~\ref{sec:asymptotics}, we present the implementation of the perturbation method to derive the conduit equation.  Information about its higher order corrections from the full set of nondimensional equations is provided in \S~\ref{sec:robustness}.  The manuscript is concluded with discussion of physical implications and future directions in \S~\ref{sec:discussion}.


\section{Problem Description} \label{sec:problem}

\begin{figure}  \centering 
	\includegraphics{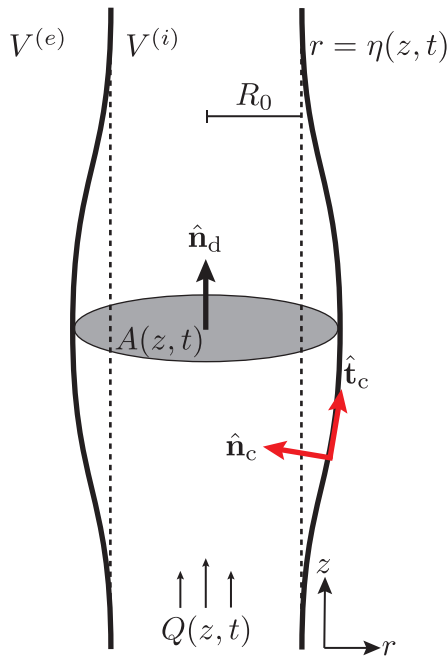}
	\caption{Geometry of an axisymmetric, intrusive fluid conduit occupying a domain $V\upi$ flowing within a dense, exterior fluid of higher viscosity in a domain $V\upe$.  The dashed line denotes an unperturbed conduit of radius $R_0$, and the solid line is the conduit with some disturbance.  Relevant physical quantities are noted, as well as the sign conventions for the conduit unit normal vector, $\nhat_\uc$, conduit unit tangent vector, $\that_\uc$, and cross-sectional disk unit normal vector, $\nhat_\ud$.}
    	\label{fig:schematic}
\end{figure}

The purpose of this section is to describe the mathematical formulation of wave propagation along an established, vertically uniform conduit, illustrated schematically in Fig.~\ref{fig:schematic}.  We consider an intrusive conduit fluid of density $\rho\upi$ and viscosity $\mu\upi$ in an extended domain denoted $V\upi$, propagating within an exterior fluid with density $\rho\upe$, viscosity $\mu\upe$, and domain $V\upe$.  The present analysis is primarily concerned with capturing the interactions in an axially symmetric conduit between buoyant forcing of the intrusive fluid due to unsteady injection and viscous stress at the conduit boundary due to the exterior fluid.  This requires the basic assumption that the intrusive fluid is less dense, $\rho\upi < \rho\upe$, and that the viscosity of the intrusive fluid is much smaller than for the exterior fluid, i.e. $\eps \ll 1$.
We will adopt the convention that $\nhat_\uc$ represents the inward pointing unit normal to the conduit surface.  The unit tangent to the conduit $\that_\uc$ is oriented so that it always has a positive vertical component.  It is also necessary to define the unit normal of cross-sectional disks along the conduit, denoted $\nhat_\ud$, oriented upward (see Fig.~\ref{fig:schematic}).

Before proceeding, we present a brief word on the notation used in this work.  We use the
cylindrical coordinate system, assuming azimuthal symmetry $(r,z)$ (see
Fig.~\ref{fig:schematic}).  Bold symbols correspond to vectors or
tensors while non-bold symbols indicate scalar quantities.  The
only superscripts we use ($\cdot\upi$, $\cdot\upe$) denote quantities
associated with the intrusive or exterior fluid, respectively.
Subscripts of non-bold symbols correspond either to the coefficients
of the velocity basis in cylindrical coordinates, e.g.
\begin{equation}
  \label{eq:5}
  \boldt{u}^{(i,e)} = 
  \begin{bmatrix}
    u_r^{(i,e)} \\ u_z^{(i,e)}
  \end{bmatrix}\ ,
\end{equation}
or numerical ordering in an asymptotic sequence.  Note that due to
axisymmetry, we only consider two vector components.  A jump in the
fluid quantity $\beta$ is denoted
\begin{equation}
  \label{eq:3}
  [\beta]_\uj = \beta\upe - \beta\upi\ ,
\end{equation}
evaluated at the fluid-fluid interface.

\subsection{Governing Equations}

The complete description of the interaction between two incompressible fluids of constant densities is given by the continuity equation for mass conservation, coupled with the Navier-Stokes equation for linear momentum conservation, which can be written compactly as
\begin{subequations}
	\label{eq:NS}
	\begin{equation}
		\label{eq:continuity}
		\del \cdot \boldt{u}^{(i,e)} = 0 \ ,
	\end{equation}	
	\begin{equation}
		\label{eq:momentum}
		\rho^{(i,e)} \ddt{\boldt{u}^{(i,e)}} = -\del p^{(i,e)} + \del \cdot \bolds{\sigma}^{(i,e)} \ .
	\end{equation}
\end{subequations}
We will utilize cylindrical coordinates assuming azimuthal symmetry and that the force due to gravity is the only external force.  Then $\boldt{u}^{(i,e)} = \left( u^{(i,e)}_r, u^{(i,e)}_z \right)$ represents the velocity vectors for the two fluids, $\mathrm{d}/\mathrm{d}t$ is the material derivative operator, and $\bolds{\sigma}^{(i,e)} = \mu^{(i,e)} \left[ \del \boldt{u} + \left( \del \boldt{u} \right)^T \right]$ is the deviatoric stress tensor.  The modified pressure $p^{(i,e)}$, or the pressure deviation from hydrostatic, can be written as the sum $p^{(i,e)} = \rho^{(i,e)} g z + P^{(i,e)}$, where $P^{(i,e)}$ is the absolute pressure.  We initially assume that the conduit has infinite vertical extent and the external fluid extends infinitely in the radial direction.

Along the center axis $r=0$, symmetry and perturbation of a vertically uniform conduit lead us to require that the intrusive fluid satisfy
\begin{equation}
	\pdr{u\upi_z} = u\upi_r = \pdr{p\upi} = 0 \ , \quad r = 0 \ .
	\label{eq:r0}
\end{equation}
In the far-field limit $r\rightarrow \infty$, we require the exterior fluid velocity field $\boldt{u}\upe$ and (modified) pressure $p\upe$ decay to zero.  The boundary between the two fluids is represented by the level curves $r = \eta(z,t)$.  The conduit area is therefore
\begin{equation}
  \label{eq:4}
  A(z,t) = \pi \eta^2(z,t) \ .
\end{equation}
The boundary is treated as a material surface, which requires the kinematic condition
\begin{equation}
	u\upi_r = \pdt{\eta} + u\upi_z \pdz{\eta} \ , \quad r = \eta(z,t) \ ,
	\label{eq:kinematic}
\end{equation}
in addition to continuity of normal and tangential components of the velocity
\begin{equation}
	\left[ \boldt{u} \cdot \nhat_\uc \right]_\uj = \left[ \boldt{u} \cdot \that_\uc \right]_\uj = 0 \ , \quad r = \eta(z,t) \ .
	\label{eq:vel_cont}
\end{equation}
The last conditions along the fluid-fluid interface which must be satisfied are the balance of normal and tangential stresses, which for general settings are written
\begin{subequations}
	\label{eq:stress_match}
	\begin{equation}
		\left[ \nhat_\uc \cdot \boldt{T} \cdot \nhat_\uc \right]_\uj = - \left( \del \cdot \nhat_\uc \right) \gamma \ , \quad r = \eta(z,t) \ ,
		\label{eq:norm_stress}
	\end{equation}
	\begin{equation}
		\left[ \that_\uc \cdot \boldt{T} \cdot \nhat_\uc \right]_\uj = \del_s \gamma \cdot \that_\uc \ , \quad r = \eta(z,t) \ ,
		\label{eq:tan_stress}
	\end{equation}	
\end{subequations} 
where $\boldt{T}^{(i,e)} = -P^{(i,e)} \boldt{I} + \bolds{\sigma}^{(i,e)}$ is the stress tensor, $\boldt{I}$ is the identity operator, $\gamma$ is the surface tension, and $\del_s = \left( \boldt{I} - \nhat_\uc \otimes \nhat_\uc \right) \cdot \del$ is the surface gradient operator for the interface.  For a more detailed description of the fluid equations and boundary conditions, see e.g. \citet{batchelor_introduction_1967}.

Since the conduit equation \eqref{eq:magma} is a PDE for the cross-sectional area, it is also convenient to derive an expression which relates the fluid properties to $A(z,t)$ \citep{scott_observations_1986,olson_solitary_1986}.  First, the volumetric flow rate $Q$ through vertical cross-sections of the conduit (see Fig.~\ref{fig:schematic}) is defined to be 
\begin{equation}
	\label{eq:flux_defn}
	Q(z,t) =  2\pi \int_{0}^{\eta(z,t)} \boldt{u}\upi \cdot \nhat_\ud \, r\, \ud r \ .
\end{equation}
Utilizing the conduit geometry, one can write the integral form of the continuity equation for cross-sectional disks as a limit of vertical cylinders of vanishing height $\delta z$ as
\begin{equation}
	\label{eq;area_cons1}
	\begin{split}
      \frac{\partial A}{\partial t} + \lim_{\delta z \rightarrow 0}  \frac{2\pi}{\delta z} &\left(  \int_0^{\eta(z+\delta z,t)}  \boldt{u}\upi \cdot \nhat_\ud \, r\, \ud r \right.  \\
	& \quad  \left. - \int_0^{\eta(z,t)} \boldt{u}\upi \cdot \nhat_\ud \,r\, \ud r \right)  = 0 \ .
	\end{split} 
\end{equation}
This can be compactly written as
\begin{equation}
	\label{eq:area_cons}
	\pdt{A} + \pdz{Q} = 0 \ .
\end{equation}
Thus the cross-sectional area of the conduit is related to the fluid properties through the intrusive fluid velocity, in particular the vertical velocity $u_z\upi$ for an axisymmetric conduit.


\subsection{Uniform Pipe Flow and Nondimensionalization}

The equations of motion will now be nondimensionalized about a state of steady, vertically uniform, conduit flow, which takes the form of a generalized Pouseuille (pipe) flow in the limit of small $\rey$ and negligible surface tension \cite{whitehead_dynamics_1975}.  In this setting, the intrusive equations are reduced to $u\upi_r = 0$, and the vertical velocity is driven by a vertical pressure gradient according to eq.~\eqref{eq:momentum}
\begin{equation}
	\label{eq:pipe}
	\frac{1}{\mu\upi} \pdz{p\upi} = \frac{1}{r} \pdr{ \ } \left(r \pdr{u\upi_z} \right) \ ,
\end{equation}
where $p\upi$ is a function of $z$ alone and the interface is fixed at $r=R_0$.  On the exterior, the velocities are small relative to the pressure, which leaves only hydrostatic pressure balance, i.e. $P\upe = -\rho\upe g z + p_0$, where $p_0$ is a constant reference pressure.  Note that these relative scaling arguments will be made more precise in \S~\ref{sec:asymptotics}, as here we seek only a background scale about which to perturb.  Imposing the boundary condition \eqref{eq:norm_stress} and neglecting surface tension effects yields an expression for the intrusive pressure
\begin{equation}
	\label{eq:buoyancy}
	p\upi = \left( \rho\upi - \rho\upe \right) g z + p_0 \ .
\end{equation}
Substituting the pressure \eqref{eq:buoyancy} into the velocity equation \eqref{eq:pipe}, performing two integrations, and imposing the boundary conditions \eqref{eq:r0} and \eqref{eq:vel_cont}, leads to the generalized pipe flow velocity for the intrusive fluid
\begin{equation}
	\label{eq:pipe_velocity}
	u\upi_z = \frac{g}{4\mu\upi} \left( \rho\upe - \rho\upi \right) \left( R_0^2 - r^2 \right) \ .
\end{equation}
Upon substitution into the tangential stress balance condition \eqref{eq:tan_stress} and neglecting surface tension, we find that the exterior fluid must have a vertical velocity which satisfies
\begin{equation}
	\label{eq:ext_velocity}
	\left. \pdr{u\upe_z} = \frac{\mu\upi}{\mu\upe} \pdr{u\upi_z} \right|_{r=R_0} \ ,
\end{equation}
and thus is proportional to the small parameter $\epsilon$.

In section \ref{sec:asymptotics}, we will perturb the interface of the steadily flowing conduit, so eq.~\eqref{eq:pipe_velocity} provides natural length and velocity scales of interest.  Previous studies have invoked a small slope assumption on conduit perturbations \cite{scott_observations_1986,olson_solitary_1986,helfrich_solitary_1990} but have not explicitly identified the appropriate vertical length scale.  We now provide the precise scaling and nondimensionalization for the governing equations that will lead to a maximal balance between buoyant and viscous stress effects.  The radial length scale $L$ is proportional to the uniform conduit radius while vertical variations are assumed to be weak according to
\begin{equation}
	\label{eq:length_scales}
	\tilde{r} = r/L \ , \quad  \tilde{z} = \epsilon^{1/2}z/L \ , \quad L = R_0/\sqrt{8} \ .
\end{equation}
The proportionality constant in the characteristic length $L$ is chosen for convenience in working with the governing equations but will be rescaled to arrive at the standard form of the conduit equation \eqref{eq:magma}.  The boundary is
now denoted by $r = \eta(z,t) = R_0 + R'(z,t)$, or $\tilde{r} = (R_0 +
R'(z,t))/L \equiv \tilde{R}(\tilde{z},\tilde{t})$.  Hence the unit
normal and tangent vectors for the conduit are given by
\begin{subequations}
	\label{eq:normal_tangential}
	\begin{equation}
		\hat{\tilde{\boldt{n}}}_\uc = \frac{1}{\|\tilde{\boldt{n}}_\uc\|} \begin{bmatrix} -1 \\ \epsilon^{1/2} \pdd{\tilde{R}}{\tilde{z}} \end{bmatrix} \ , \quad
		\hat{\tilde{\boldt{t}}}_\uc = \frac{1}{\|\tilde{\boldt{t}}_\uc\|} \begin{bmatrix} \epsilon^{1/2} \pdd{\tilde{R}}{\tilde{z}} \\ 1 \end{bmatrix}	\ ,
	\end{equation}
\end{subequations}
where
\begin{equation}
	\|\tilde{\boldt{n}}_\uc\| = \|\tilde{\boldt{t}}_\uc\| = \left[  1 + \epsilon \left(\pdd{\tilde{R}}{\tilde{z}} \right)^2 \right]^{-1/2} \ .
\end{equation}
Velocities are normalized to the radially-averaged vertical velocity of the uniform conduit 
\begin{equation}
	\label{eq:velocity_scale}
	\tilde{\boldt{u}}^{(i,e)} = \boldt{u}^{(i,e)} / U  \ , \quad U = \frac{g R_0^2 \left( \rho\upe - \rho\upi \right)}{8\mu\upi} \ ,
\end{equation}
leading to the long time scale $\epsilon^{-1/2}T$ for vertical dynamics where
\begin{equation}
	\label{eq:time_scale}
	\tilde{t} = \eps^{1/2}t/T \ , \quad T = L/U \ .
\end{equation}
To nondimensionalize the pressure, the characteristic scale $\Pi$ is chosen so that the vertical pressure gradient within the conduit balances the viscous force due to radial variation in the vertical velocity, 
\begin{equation}
	\label{eq:pressure_scale}
\tilde{p}^{(i,e)} = 	\eps^{1/2} \frac{p^{(i,e)} - p_0}{\Pi}  \ , \quad \Pi = \mu\upi U/L \ .
\end{equation}
Like in dimensional variables, the nondimensional, modified pressure can be decomposed as $\tilde{p}^{(i,e)} =  \tilde{P}^{(i,e)}-\tilde{p}_h^{(i,e)}$, where $\tilde{P}^{(i,e)} = \eps^{1/2}P^{(i,e)}/\Pi$ is the scaled, absolute pressure and $\tilde{p}^{(i,e)}_h$ is the normalized hydrostatic pressure which takes the form
\begin{equation}
	\label{eq:rescaled_hydrostatic}
	\tilde{p}^{(i,e)}_h = -\eps^{1/2} \frac{\rho^{(i,e)}gz}{\Pi} = \frac{-\rho^{(i,e)}\tilde{z}}{\rho\upe-\rho\upi}\   \ .
\end{equation}
Surface tension was neglected in the discussion of the uniform conduit, but it will be included in the full system of equations for completeness, so it is normalized about a characteristic scale $\Gamma$:
\begin{equation}
	\label{eq:sfc_tension_scale}
	\tilde{\gamma} = \gamma / \Gamma \ .
\end{equation}
The Reynolds numbers for the viscous fluid conduit system are therefore defined for the two fluids according to
\begin{equation}
	\label{eq:reynolds}
	\rey^{(i,e)} = \frac{\rho^{(i,e)}\left(\eps^{-1/2} L\right) U}{\mu^{(i,e)}}  \ .
\end{equation}
Note that $\rey\upe = \eps \left( \rho\upe/\rho\upi \right) \rey\upi$.  While $\rho\upi < \rho\upe$, we typically consider fluids where $\rey\upe \ll \rey\upi$.

The governing equations in nondimensional form are obtained by direct substitution of the above scalings into the set of dimensional equations and boundary conditions.  For the intrusive fluid, the nondimensional continuity equation \eqref{eq:continuity} is 
\begin{equation}
	\label{eq:int_cont}
	\frac{1}{\tilde{r}} \pdd{ \ }{\tilde{r}} \left( \tilde{r} \tilde{u}\upi_{\tilde{r}} \right)+ \eps^{1/2} \pdd{\tilde{u}\upi_{\tilde{z}}}{\tilde{z}} = 0 \ ,
\end{equation}
and linear momentum balance \eqref{eq:momentum} requires
\begin{subequations}
	\label{eq:int_linear_mom}
	\begin{equation}
		\label{eq:int_rad_mom}
		\rey\upi \did{\tilde{u}\upi_{\tilde{r}}}{\tilde{t}} = -\eps^{-3/2}\pdd{\tilde{p}\upi}{\tilde{r}} + \tilde{\del}^2 \tilde{u}\upi_{\tilde{r}} - \eps^{-1} \frac{\tilde{u}\upi_{\tilde{r}}}{\tilde{r}^2} \ ,
	\end{equation}
	\begin{equation}
		\label{eq:int_vert_mom}
		\rey\upi \did{\tilde{u}\upi_{\tilde{z}}}{\tilde{t}} = -\eps^{-1} \pdd{\tilde{p}\upi}{\tilde{z}} + \tilde{\del}^2 \tilde{u}\upi_{\tilde{z}} \ ,
	\end{equation}
\end{subequations}
where 
\begin{equation}
	\label{eq:total_deriv_nondim}
	\frac{\ud^{(i,e)} \ }{\ud \tilde{t}} = \pdd{ \ }{\tilde{t}} + \left( \tilde{\boldt{u}}^{(i,e)} \cdot \tilde{\del} \right) \ ,
\end{equation}
and
\begin{equation}
	\label{eq:del_nondim}
	\tilde{\del} = \eps^{-1/2} \pdd{ \ }{\tilde{r}} + \pdd{ \ }{\tilde{z}} \ , \quad
	\tilde{\del}^2 = \eps^{-1} \frac{1}{\tilde{r}} \pdd{ \ }{\tilde{r}} \left( \tilde{r} \pdd{ \ }{\tilde{r}} \right) + \frac{\partial^2 \ }{\partial \tilde{z}^2} \ .
\end{equation}
For the exterior fluid, mass conservation \eqref{eq:continuity} is similarly
\begin{equation}
	\label{eq:ext_cont}
	\frac{1}{\tilde{r}} \pdd{ \ }{\tilde{r}} \left( \tilde{r} \tilde{u}\upe_{\tilde{r}} \right)+ \eps^{1/2} \pdd{\tilde{u}\upe_{\tilde{z}}}{\tilde{z}} = 0 \ ,
\end{equation}
and momentum balance \eqref{eq:momentum} requires
\begin{subequations}
	\label{eq:ext_linear_mom}
	\begin{equation}
		\label{eq:ext_rad_mom}
		\rey\upe \did{\tilde{u}\upe_{\tilde{r}}}{\tilde{t}} = -\eps^{-1/2}\pdd{\tilde{p}\upe}{\tilde{r}} + \tilde{\del}^2 \tilde{u}\upe_{\tilde{r}} - \eps^{-1} \frac{\tilde{u}\upe_{\tilde{r}}}{\tilde{r}^2} \ ,
	\end{equation}
	\begin{equation}
		\label{eq:ext_vert_mom}
		\rey\upe \did{\tilde{u}\upe_{\tilde{z}}}{\tilde{t}} = - \pdd{\tilde{p}\upe}{\tilde{z}} + \tilde{\del}^2 \tilde{u}\upe_{\tilde{z}} \ .
	\end{equation}
\end{subequations}

The boundary conditions along the axis of symmetry $\tilde{r}=0$, given by eq.~\eqref{eq:r0}, are now
\begin{equation}
	\label{eq:r0_nondim}
	\pdd{\tilde{u}\upi_{\tilde{z}}}{\tilde{r}} = 0 \ , \quad \tilde{u}\upi_{\tilde{r}} = 0 \ , \quad \pdd{\tilde{p}\upi}{\tilde{r}} = 0 \ ,\quad \tilde{r}=0 \ ,
\end{equation}
and the far-field boundary condition still requires that the exterior fluid velocities and (modified) pressure decay as $\tilde{r}\rightarrow \infty$.  The kinematic boundary condition \eqref{eq:kinematic} becomes
\begin{equation}
	\label{eq:kinematic_nondim}
	\tilde{u}\upi_{\tilde{r}} =  \eps^{1/2} \left( \pdd{\tilde{R}}{\tilde{t}} + \tilde{u}\upi_{\tilde{z}} \pdd{\tilde{R}}{\tilde{z}} \right) \ , \quad \tilde{r} = \tilde{R}(\tilde{z},\tilde{t}) \ ,
\end{equation}
and couples to the equations for continuity of the velocity components \eqref{eq:vel_cont}, which are now 
\begin{subequations}
	\label{eq:vel_cont_nondim}
	\begin{equation}
		\label{eq:norm_vel}
		\left( \tilde{u}\upi_{\tilde{r}} - \tilde{u}\upe_{\tilde{r}} \right) = \eps^{1/2} \pdd{\tilde{R}}{\tilde{z}} \left( \tilde{u}\upi_{\tilde{z}} - \tilde{u}\upe_{\tilde{z}} \right) \ , 
	\end{equation}
	\begin{equation}
		\label{eq:tan_vel}
		\left( \tilde{u}\upe_{\tilde{z}} - \tilde{u}\upi_{\tilde{z}} \right) = \eps^{1/2} \pdd{\tilde{R}}{\tilde{z}} \left( \tilde{u}\upi_{\tilde{r}} - \tilde{u}\upe_{\tilde{r}} \right) \ .
	\end{equation}
\end{subequations}
To nondimensionalize the stress balance conditions \eqref{eq:stress_match}, the normalized deviatoric stress tensor $\tilde{\bolds{\sigma}}^{(i,e)} = \eps^{1/2}(L/\mu^{(i,e)}U) \bolds{\sigma}^{(i,e)}$ can be written 
\begin{equation}
	\label{eq:stress_nondim}
	\begin{split}
	\tilde{\bolds{\sigma}}^{(i,e)} & = 
	\begin{bmatrix} \tilde{\sigma}^{(i,e)}_{\tilde{r}\tilde{r}} & \tilde{\sigma}^{(i,e)}_{\tilde{r}\tilde{z}} \\ \tilde{\sigma}^{(i,e)}_{\tilde{z}\tilde{r}} & \tilde{\sigma}^{(i,e)}_{\tilde{z}\tilde{z}} \end{bmatrix} \\
	& = \eps^{1/2} 
	\begin{bmatrix}
	2 \pdd{\tilde{u}^{(i,e)}_{\tilde{r}}}{\tilde{r}} &  \pdd{\tilde{u}^{(i,e)}_{\tilde{z}}}{\tilde{r}} + \eps^{1/2} \pdd{\tilde{u}^{(i,e)}_{\tilde{r}}}{\tilde{z}}  \\
	\pdd{\tilde{u}^{(i,e)}_{\tilde{z}}}{\tilde{r}} + \eps^{1/2} \pdd{\tilde{u}^{(i,e)}_{\tilde{r}}}{\tilde{z}}  & 2 \eps^{1/2} \pdd{\tilde{u}^{(i,e)}_{\tilde{z}}}{\tilde{z}} 
	 \end{bmatrix} \ .
	 \end{split}
\end{equation}
Then the nondimensional stress balance conditions in the normal and tangential directions become, respectively,
\begin{subequations}
	\label{eq:stress_match_nondim}
	\begin{equation}
		\label{eq:normal_stress_nondim}
		\begin{split}
			& \left[
			-\|\tilde{\boldt{n}}_\uc\|^2 \tilde{P} + \kappa \left( \tilde{\sigma}_{\tilde{r}\tilde{r}} - 2\eps^{1/2} \pdd{\tilde{R}}{\tilde{z}} \tilde{\sigma}_{\tilde{r}\tilde{z}} + 
			\eps \left( \pdd{\tilde{R}}{\tilde{z}} \right)^2  \tilde{\sigma}_{\tilde{z}\tilde{z}} \right) \right]_\uj \\
			&  \quad = \left. \frac{\eps^{1/2} \|\tilde{\boldt{n}}_\uc\|}{\mathrm{Ca}} \left( \frac{1}{\tilde{R}} - \eps \frac{\partial^2 \tilde{R}}{\partial \tilde{z}^2} \right) \tilde{\gamma}
			\ \right|_{\tilde{r} = \tilde{R}(z,t)} \ ,
		\end{split}
	\end{equation}
	\begin{equation}
		\label{eq:tan_stress_nondim}
		\begin{split}
			& \left[ \kappa \left\{
			- \left( 1 - \eps \left( \pdd{\tilde{R}}{\tilde{z}} \right)^2 \right) \tilde{\sigma}_{\tilde{r}\tilde{z}} + \eps^{1/2} \pdd{\tilde{R}}{\tilde{z}} \left( \tilde{\sigma}_{\tilde{z}\tilde{z}} - \tilde{\sigma}_{\tilde{r}\tilde{r}} \right) 
			\right\} \right]_\uj \\
			& \quad \left. = \frac{\eps \|\tilde{\boldt{n}}_\uc\|}{\mathrm{Ca}} \left( \pdd{ \ }{\tilde{z}} + \pdd{\tilde{R}}{\tilde{z}} \pdd{ \ }{\tilde{r}} \right) \tilde{\gamma} \ 
			\right|_{\tilde{r} = \tilde{R}(z,t)} \ ,
		\end{split}
	\end{equation}
\end{subequations}
where $\kappa$ is a fluid-specific coefficient such that $\kappa^{(e)} = \eps^{-1}, \ \kappa^{(i)} = 1$ and the capillary number is defined to be $\mathrm{Ca} = \mu\upi U/\Gamma$.

This is now a complete system of nondimensional equations with boundary conditions.  It is important to note that the system of equations to this point is general, with no approximations.


\section{Derivation of the Approximate Model} \label{sec:asymptotics}

Using the normalizations in the previous section and now treating the dimensionless parameters $\eps$, $\eps \rey\upi$, and $1/\mathrm{Ca}$ as small in comparison to unity, the leading order behavior of the nondimensional model equations will be derived.  We determine the appropriate scalings of the fluid quantities resulting in the long-time validity of a maximal balance between buoyancy and viscous stress and, ultimately, the conduit equation \eqref{eq:magma}.  By our choice of scaling, the radial dynamics of the conduit fluid are captured, as well as the near interface dynamics in the exterior.  It is important to consider that because the radial extent is large, the expressions derived below are valid only near the boundary and do not represent a uniform asymptotic expansion across large radial distances.

Note that the tilde notation for nondimensional variables will be dropped for the remainder of the manuscript, but all quantities are understood to be nondimensional.


\subsection{Asymptotic Expansions and Scaling Ansatz} \label{sec:scale}

The velocities and pressures of the two fluids, as well as the dimenionless parameters, will now be scaled and expanded in powers of $\eps$.  

The dimensional equations were normalized so that vertical pressure gradients in the intrusive fluid balance with viscous forces due to the vertical flow, which implies $p\upi, u\upi_z = \mathcal{O}(1)$ and can be expanded
\begin{equation}
	\label{eq:pressure_int}
	p\upi = p\upi_0 + \eps p\upi_1 + \cdots \ , 
\end{equation}
\begin{equation}
	\label{eq:vert_int}
	u\upi_z = u\upi_{z,0} + \eps u\upi_{z,1} + \cdots \ .
\end{equation}
This also requires that the inertial terms in the interior be neglected.  From \eqref{eq:int_vert_mom}, this means $\rey\upi \ll \eps^{-1}$.  Note that this implies $\rey\upe \ll \rho\upe/\rho\upi$ so typically $\rey\upe \ll 1$.  The length scales \eqref{eq:length_scales} were then chosen so that the uniform conduit radius $R_0$ is $\mathcal{O}(1)$ and the vertical dynamics occur on a longer spatial scale proportional to $\eps^{-1/2}$.  The small slope condition therefore requires $R \ll \eps^{-1/2}$.  We emphasize that the deviation of the conduit wall from uniformity can be very large.

In the uniform conduit, the intrusive modified pressure was balanced with the hydrostatic pressure of the exterior fluid, but the exterior modified pressure was negligible.  In the present scaling, the extrusive pressure can be represented by
\begin{equation}
	\label{eq:pressure_ext}
	p\upe = \eps \left( p\upe_1 + \eps p\upe_2 + \cdots \right) \ ,
\end{equation}
so that leading order hydrostatic balance, $p\upe_h = P\upe_0$, is preserved.  Tangential shear stress due to the intrusive vertical velocity was balanced by an exterior vertical velocity with scale set by eq.~\eqref{eq:ext_velocity} so that according to \eqref{eq:vert_int},
\begin{equation}
	\label{eq:vert_ext}
	u\upe_z = \eps \left( u\upe_{z,0} + \eps u\upe_{z,1} + \cdots \right) \ .
\end{equation}
The use of miscible fluids renders the surface tension negligible, which by \eqref{eq:stress_match_nondim} means $1/\mathrm{Ca} \ll 1$.  Lastly, the amplitudes of the radial velocities are identified from \eqref{eq:kinematic_nondim} and expanded asymptotically as
\begin{equation}
	\label{eq:rad_int}
	u\upi_r = \eps^{1/2} \left( u\upi_{r,0} + \eps u\upi_{r,1} + \cdots \right) \ ,
\end{equation}
\begin{equation}
	\label{eq:rad_ext}
	u\upe_r = \eps^{1/2} \left( u\upe_{r,0} + \eps u\upe_{r,1} + \cdots \right) \ .
\end{equation}
The fluid quantities have now been scaled so that disturbances along the uniform conduit are captured by appealing only to the leading order, approximate governing equations.  In previous derivations, the key physical assumption made was that the vertical wavelength of conduit perturbations was long in comparison with the radial amplitude.  In our framework, this is captured by normalizing to the radial, conduit scale and then allowing slow variations in the vertical direction.  The magnitude of the slope is proportional to the ratio of the intrusive radial velocity to vertical velocity, which is $\mathcal{O}(\eps^{1/2})$ and thus indeed small in our analysis.


\subsection{Leading Order Solution} \label{sec:lead}

In the physical regime of validity of the rescaled fluid properties, the leading order nondimensional equations for the intrusive fluid are
\begin{subequations}
	\label{eq:lead_int}
	\begin{equation}
		\label{eq:lead_continuity_int}
		\frac{1}{r} \pdr{ \ } \left(r u\upi_{r,0} \right)+ \pdz{u\upi_{z,0}} = 0 \ ,
	\end{equation}
	\begin{equation}
		\label{eq:lead_rad_mom_int}
		\pdr{p\upi_0} = 0 \ ,
	\end{equation}
	\begin{equation}
		\label{eq:lead_vert_mom_int}
		\pdz{p\upi_0} = \frac{1}{r} \pdr{ \ } \left( r \pdr{u\upi_{z,0}} \right) \ ,
	\end{equation}
\end{subequations}
and for the extrusive fluid
\begin{subequations}
	\label{eq:lead_ext}
	\begin{equation}
		\label{eq:lead_continuity_ext}
		\frac{1}{r} \pdr{ \ } \left(r u\upe_{r,0} \right) = 0 \ ,
	\end{equation}
	\begin{equation}
		\label{eq:lead_rad_mom_ext}
		\frac{1}{r} \pdr{ \ } \left( r \pdr{u\upe_{r,0}} \right) - \frac{u\upe_{r,0}}{r^2} = 0 \ ,
	\end{equation}
	\begin{equation}
		\label{eq:lead_vert_mom_ext}
		\frac{1}{r} \pdr{ \ } \left( r \pdr{u\upe_{z,0}} \right) = 0 \ .
	\end{equation}
\end{subequations}
The axial boundary conditions at $r=0$ \eqref{eq:r0_nondim} and the far-field conditions remain unchanged.  Along the interface, the velocity continuity conditions \eqref{eq:vel_cont_nondim} and kinematic condition \eqref{eq:kinematic_nondim} at leading order are
\begin{equation}
	\label{eq:vel_cont_lead}
	u\upi_{z,0} = 0 \ , \quad u\upi_{r,0} = u\upe_{r,0} = \pdt{R(z,t)} \ , \quad r = R(z,t) \ .
\end{equation}	
The interfacial force balance equations \eqref{eq:stress_match_nondim} simplify to
\begin{subequations}
	\label{eq:stress_match_lead}
	\begin{equation} 
		\label{eq:norm_stress_lead}
		P\upe_0 - P\upi_0 = 2\pdr{u\upe_{r,0}} \ , \quad r = R(z,t) \ , 
	\end{equation}
	\begin{equation}
		\label{eq:tan_stress_lead}
		\pdr{u\upi_{z,0}} = \pdr{u\upe_{z,0}} - \pdz{u\upe_{r,0}}+ 2\pdz{R} \pdr{u\upe_{r,0}}\ , \quad r = R(z,t) \ .
	\end{equation}
\end{subequations}

In this form, analytical expressions for the fluid properties can be found.  From \eqref{eq:lead_continuity_ext} and \eqref{eq:lead_rad_mom_ext}, along with the boundary condition \eqref{eq:vel_cont_lead}, the extrusive radial velocity is
\begin{equation}
	\label{eq:rad_ext_soln}
	u\upe_{r,0} =  \pdt{R(z,t)} \frac{R(z,t)}{r} \ ,
\end{equation}
hence exhibits algebraic decay. 
From our scaling of the modified pressure \eqref{eq:pressure_ext}, the exterior absolute pressure is
\begin{equation}
	\label{eq:pressure_ext_soln}
	P\upe_0 = - \frac{\rho\upe z}{\rho\upe - \rho\upi} \ ,
\end{equation}
and because the interior pressure is independent of $r$ \eqref{eq:lead_rad_mom_int}, the boundary condition \eqref{eq:stress_match_lead} gives the form of the intrusive pressure to be
\begin{equation}
	\label{eq:pressure_int_soln}
	p\upi_0 = P\upi_0 + \frac{\rho\upi z}{\rho\upe - \rho\upi} = \frac{2}{R(z,t)} \pdt{R(z,t)} - z \ .
\end{equation}
Substituting \eqref{eq:pressure_int_soln} into \eqref{eq:lead_vert_mom_int} and imposing the boundary conditions \eqref{eq:r0_nondim} and \eqref{eq:vel_cont_lead} yields
\begin{equation}
	\label{eq:vert_int_soln}
	u\upi_{z,0} = \frac{1}{4} \left[1 - \pdz{ \ } \left( \frac{2}{R(z,t)} \pdt{R(z,t)} \right) \right] \left( R^2(z,t) - r^2 \right) \ .
\end{equation}
The dependence of $u\upi_{r,0}$ on the interface can be derived from \eqref{eq:lead_continuity_int}, which can be rewritten
\begin{equation}
	\label{eq:cont_int_rewrite}
	u\upi_{r,0} = -\frac{1}{r} \pdz{ \ } \int_0^{r} r' u\upi_{z,0} \ \ud r' \ .
\end{equation}
From this form, it is readily observed that by imposing the kinematic boundary condition \eqref{eq:vel_cont_lead}, the resulting equation \eqref{eq:cont_int_rewrite} is identical to the area conservation law \eqref{eq:area_cons}.  While this is not surprising because the conservation law was derived by the integral form of the continuity equation within the intrusive conduit, it shows that the multiple scales approach leads directly to the appropriate relationship between the dynamic interface and the fluid velocity field, without appealing to physical intuition.  Solving \eqref{eq:cont_int_rewrite} gives the explicit form of the intrusive, radial velocity to be
\begin{equation}
	\label{eq:rad_int_soln}
	\begin{split}
	u\upi_{r,0} &= -\frac{1}{8r} \pdz{ \ } \left\{ \left[1 - \pdz{ \ } \left( \frac{2}{R(z,t)} \pdt{R(z,t)} \right) \right]  \right. \\
	& \left. \left( R^2(z,t) - \frac{r^2}{2} \right)r^2 \right\} \ .
	\end{split}
\end{equation}

The exterior vertical velocity can be solved according to \eqref{eq:lead_vert_mom_ext} along with the tangential shear stress balance condition \eqref{eq:tan_stress_lead} to obtain
\begin{equation}
	\label{eq:vert_ext_soln}
	u\upe_{z,0} = f(z,t) \ln{r} + g(z,t) \ ,
\end{equation}
where $f(z,t)$ satisfies
\begin{equation}
	\label{eq:f_coeff}
	f(z,t) = 2 \pdt{ \ } \left( R(z,t) \pdz{R(z,t)} \right) - \frac{1}{2} R^2(z,t) \ .
\end{equation}
We note here that while $u\upe_{r,0}$ decays as $r\rightarrow\infty$ and the pressure at leading order is hydrostatic as required by the far-field boundary condition, the vertical velocity $u\upe_{z,0}$ does not decay.  As discussed earlier, the leading order equations for the exterior fluid are valid only near the interface.  It is a known phenomenon that there is no solution to the Stokes' flow equations, as considered here, in cylindrical coordinates which vanish at $\infty$ (cf. \citep{batchelor_introduction_1967}).  To satisfy this boundary condition, and to solve for the undetermined function $g(z,t)$, one must appeal to higher order terms, where inertial effects are incorporated.  We do not consider this calculation here but will discuss the solution for $u\upe_{z,0}$ when a radial boundary is present in \S \ref{sec:robustness}.

Hence the perturbed problem has subtle differences from the uniform conduit problem.  There is now a pressure jump across the interface resulting from the unsteady injection rate.  This is balanced by a viscous, normal stress from the exterior fluid.  All leading order fluid quantities have been determined in terms of the radial profile of the conduit interface $R(z,t)$.


\subsection{Reduction to the Magma Equation} \label{sec:magma}

The fluid properties are now known at leading order, but it remains to find the leading order governing equation for the interfacial dynamics.  This is done by appealing to the area conservation law \eqref{eq:area_cons} or, equivalently, by imposing the kinematic boundary condition \eqref{eq:vel_cont_lead} on the intrusive radial velocity expression \eqref{eq:cont_int_rewrite}.  Integrating the vertical velocity expression \eqref{eq:vert_int_soln} and relating the conduit radius to the area via $A(z,t) = \pi R^2(z,t)$, yields an expression for the volumetric flow rate $Q(z,t)$ within the conduit in terms of its area,
\begin{equation}
	\label{eq:flux}
	Q(z,t) = \frac{A^2}{8\pi}\left\{ 1 - \pdz{ \ } \left( A^{-1}\pdt{A} \right) \right\} \ .
\end{equation}
Rescaling the area so that $A' = A/8\pi$ and inserting \eqref{eq:flux} into the area conservation law, we are left with
\begin{equation}
	\pdt{A'} + \pdz{ \ } \left\{ A'^2 \left[ 1 -  \pdz{ \ } \left( A'^{-1}\pdt{A'} \right) \right] \right\} = 0 \ ,
\end{equation}
which is precisely the conduit equation $\eqref{eq:magma}$ in standard form.  


\section{Robustness of the Magma Equation} \label{sec:robustness}

The benefit of the multiple scales approach to the derivation of eq.~\eqref{eq:magma} is that the order of magnitude of higher order corrections and the relation between dimensionless parameters are elucidated.  In particular, we are interested in criteria which indicate when previously neglected terms become important to the leading order interfacial dynamics.  This section is devoted to exploring the effects of several assumptions made in \S~\ref{sec:asymptotics} and identifying approximate points of breakdown for the conduit equation.  We also demonstrate the ability to derive information about higher order corrections in special cases.  In what follows, we determine the scalings such that all corrections to the conduit equation \eqref{eq:magma} are $\mathcal{O}(\eps)$.

\subsection{Viscous, Higher Order Corrections} \label{sec:viscous}

The equations solved in deriving the conduit equation \eqref{eq:magma} in \S~\ref{sec:lead} were a special case of the Stokes' flow equations, in which the vertical dynamics occured over a much longer length scale than the radial dynamics.  A convenient analytical property of the axisymmetric Stokes' flow equations, is that one can rewrite the nondimensional equations in the form \citep{payne_stokes_1960}
\begin{equation}
	\label{eq:harmonic}
	\eps \tilde{\del}^2 p^{(i,e)} = \frac{1}{r} \pdr{ \ } \left(r \pdr{p^{(i,e)}} \right) + \eps \frac{\partial^2 p^{(i,e)}}{\partial z^2} = 0 \ 
\end{equation}
\begin{equation}
	\label{eq:streamfun}
 	\tilde{\mathcal{L}}^2 \psi^{(i,e)} = 0 \ , \quad \tilde{\mathcal{L}} = \eps \frac{\partial^2 \ }{\partial z^2} + \frac{\partial^2 \ }{\partial r^2} - \frac{1}{r} \pdr{ \ } 
\end{equation}
where $\psi^{(i,e)}$ is the Stokes' stream function, which is related to the velocity components by
\begin{equation}
	u^{(i,e)}_r = -\eps^{1/2}\frac{1}{r} \pdz{\psi^{(i,e)}} \ , \quad u^{(i,e)}_z = \frac{1}{r} \pdr{\psi^{(i,e)}} \ .
\end{equation}
In the asymptotic formulation, the fluid pressures and velocities were expanded in asymptotic series and expressions for the leading order term in the expansion were found.  It was unclear from the form of the equations, what would be the order of magnitude of the second term in the series, but expanding \eqref{eq:harmonic}, \eqref{eq:streamfun}, one can see that $p^{(i,e)}_1 = \mathcal{O}(\eps p^{(i,e)}_0)$ and $\boldt{u}^{(i,e)}_1 = \mathcal{O}(\eps \boldt{u}^{(i,e)}_0)$, provided the Stokes' regime is still valid.  Otherwise, inertial effects require an alternative scaling.  Because the conduit equation \eqref{eq:magma} was derived from the intrusive vertical velocity, this implies that viscous corrections to the conduit equation \eqref{eq:magma} will be of magnitude $\mathcal{O}(\eps)$.

\subsection{Breakdown of Assumptions} \label{sec:breakdown}
A key physical assumption in deriving this model equation is the choice of characteristic scales about which the full governing equations were normalized.  In choosing the length scales \eqref{eq:length_scales} and imposing $\eps \ll 1$, we implicitly assumed, by dropping higher order effects, that the (dimensionless) radial dynamics occur such that $r \ll \eps^{-1/2}$.  Specifically, if the perturbed conduit radius satisfies $\eps^{1/2}R(z,t) \ll 1$, then our assumptions of dominant balance along the interface are preserved, as well as the characteristic scalings chosen.  However, when this condition is violated, our asymptotic construction fails.

A similar condition can be derived for the breakdown of the assumption to neglect inertial effects.  From eq.~\eqref{eq:int_vert_mom}, if the interior Reynolds number satisfies $\rey\upi = \mathcal{O}(1)$, then inertial corrections to the conduit equation \eqref{eq:magma} will have magnitude $\eps$.  This condition is equivalent to the small slope condition, which can be seen by direct evaluation of \eqref{eq:reynolds}.  Hence violation of the small slope assumption coincides with the introduction of inertial effects to the viscous conduit.

Another key assumption made in deriving the conduit equation was to neglect surface tension effects.  In the case of miscible fluids, this is reasonable, but the conduit equation is valid more generally, with surface tension effects entering the approximate model \eqref{eq:magma} at $\mathcal{O}(\eps)$, provided the capillary number satisfies $\mathrm{Ca} = \mathcal{O}(\eps^{-1/2})$.

\subsection{Dependence on an Outer Wall}

Suppose that instead of considering the exterior dynamics on the infinite half-line, $r>0$, with far-field boundary conditions, we confined the exterior fluid with an outer wall, say at $r=L_{\mathrm{w}} \gg R_0$, and imposed a no-slip, no-penetration boundary condition, i.e. both components of the exterior velocity field are zero.  The goal is to understand how the value of $L_{\mathrm{w}}$ couples into the approximate model.  Assuming the same scalings as in \S~\ref{sec:asymptotics}, the change comes in the solution to the exterior velocities.  To prevent confusion with earlier work, we will use $\boldt{v}^{(i,e)}$ to indicate velocities in this section.  Using eq.~\eqref{eq:lead_rad_mom_ext}, and imposing the boundary conditions at the outer wall, the particular form of the radial velocity is now
\begin{equation} 
	v\upe_{r,0} = \frac{R(z,t)}{R^2(z,t)-L_{\mathrm{w}}^2} \pdt{R(z,t)} \left( r -L_{\mathrm{w}}^2 r^{-1} \right) \ .
\end{equation}
By eq.~\eqref{eq:vert_ext_soln} and imposing the no-slip condition at $L_{\mathrm{w}}$, the exterior vertical velocity is
\begin{equation}
	v\upe_{z,0} = f(z,t) \ln{\left(r/L_{\mathrm{w}}\right)} \ ,
\end{equation}
where $f(z,t)$ is defined by eq.~\eqref{eq:f_coeff}.  Note that the inclusion of an exterior wall allows for the complete characterization of the leading order velocity field, while in the case of infinite radius, one would have to appeal to higher order to determine the free function $g(z,t)$.  The exterior mass conservation equation \eqref{eq:lead_continuity_ext} is satisfied to leading order only if $L_{\mathrm{w}}\gg 1$.  To see how $L_{\mathrm{w}}$ incorporates into the approximate model for the interface, the normal stress match condition is modified to include the new form of the velocity, so that upon substitution, it becomes 
\begin{equation}
	v\upi_{z,0} = \frac{1}{4} \left[1 + \pdz{ \ } \left( \frac{2(R^2 + L_{\mathrm{w}}^2)}{R(R^2-L_{\mathrm{w}}^2)} \pdt{R} \right) \right]\left( R^2 - r^2 \right) \ .
\end{equation}
Hence the flux in the presence of an outer wall, denoted $q(z,t)$, satisfies
\begin{equation}
	q(z,t) = \frac{A^2}{8\pi} \left\{ 1+  \pdz{ \ } \left[ A^{-1} \pdt{A} \left(\frac{A+\pi L_{\mathrm{w}}^2}{A-\pi L_{\mathrm{w}}^2} \right)  \right] \right\} \ ,
\end{equation}
which in the limit $L_{\mathrm{w}}\rightarrow \infty$, is exactly \eqref{eq:flux}.  However, if $1\ll L_{\mathrm{w}} < \infty$, then a correction to \eqref{eq:magma} can be derived.  Expanding about $1/L_{\mathrm{w}} = 0$ yields a first-order correction of the form
\begin{equation}
	q(z,t) \sim \frac{A^2}{8\pi} \left\{ 1+  \pdz{ \ } \left[ A^{-1} \pdt{A} \left( -1-2\frac{A}{\pi L_{\mathrm{w}}^2} \right)  \right] \right\} \ ,
\end{equation}
which is an additional higher order dispersive term.  Hence, the modified conduit equation becomes
\begin{equation}
	\pdt{A'} + \pdz{ \ } \left\{A'^2 \left[ 1- \pdz{ \ } \left( A'^{-1} \pdt{A'} \right) - \frac{16}{L_{\mathrm{w}}^2}\frac{\partial^2{A'}}{\partial t \ \partial z} \right]	\right\} = 0 \ .
\end{equation}
Thus, so long as $L_\mathrm{w} = \mathcal{O}(\eps^{-1/2})$, the conduit equation \eqref{eq:magma} is valid to $\mathcal{O}(\eps)$.

\subsection{Symmetry Breaking -- Inclined Conduit}

Due to the small but nonnegligble velocities of the exterior fluid, it is unlikely that the axisymmetric assumption is entirely accurate for an experimentally generated conduit.  Here we consider the point of breakdown due to an intrusive conduit fluid flowing at an angle $\phi$ relative to the vertical axis, with the biggest difference now being angular effects.  The coordinate system is now rotated by the vertical angle $\phi$ so that $z'$ is the longitudinal axis of the conduit while $r'$ is the radial direction and $\theta'$ is the angular.  The aim is to understand how the vertical angle introduces higher order corrections in the physical setting in which exterior, radial, viscous effects balance buoyant forcing within the conduit.  Therefore, we will neglect the angular momentum equation in seeking an approximate criterion, but understand that this introduces an additional possibility for breakdown.  In fact, via linear stability analysis, a critical tilt angle for gravitational instability of the intrusive fluid has been derived \citep{whitehead_instabilities_1982}, but that study did not address the validity of the approximate model \eqref{eq:magma}.  

The nondimensional pressure in the new coordinates can be written $\hat{p}^{(i,e)} = \hat{P}^{(i,e)} - \hat{p}_h^{(i,e)}$, where the hydrostatic pressure is now
\begin{equation}
	\label{eq:nondim_hydrostatic}
	\begin{split}
	\hat{p}^{(i,e)}_h &= \frac{\rho^{(i,e)}}{\rho\upe-\rho\upi} \left[
	-z'\cos{\phi} \right.\\
	&+ \left. \eps^{1/2}r'\sin{\phi} \left( \sin{\theta'} + \cos{\theta'} \right) \right] \ .
	\end{split}
\end{equation}
By continuing to assume a leading order, hydrostatic balance on the exterior, the extrusive radial velocity is unchanged, but normal stress balance \eqref{eq:norm_stress_lead} and the radial momentum equation \eqref{eq:lead_rad_mom_ext} yield the intrusive pressure to be
\begin{equation}
	\hat{p}\upi_0 = -z'\cos{\phi} + 2\left( R^{-1} \pdt{R} \right) + \eps^{1/2}R\sin{\phi} \left( \sin{\theta'}+\cos{\theta'}\right) \ .
\end{equation}
Then from \eqref{eq:vel_cont_lead}, it is seen that the $\mathcal{O}(1-\cos{\phi})$ and $\mathcal{O}\left( \eps^{1/2} \sin{\phi} \right)$ corrections to the pressure will enter at the same magnitude into the intrusive, vertical velocity, and hence into the area conservation law in the same manner.  We note, however, that for $\phi \ll 1$, we indeed recover the conduit equation.  Then provided the angular momentum balance equation holds as well, the conduit equation \eqref{eq:magma} is valid to $\mathcal{O}(\eps)$, as long as $\phi = \mathcal{O}(\eps^{1/2})$.

\subsection{Summary}

When higher order corrections are of size $\mathcal{O}(\eps)$, the conduit equation \eqref{eq:magma} is valid for times $t \ll 1/\eps$ and amplitudes $A \ll 1/\eps$, in the limit $\eps \rightarrow 0$.  The required parameter constraints discussed in this section take the form:
\begin{itemize}
	\item Weak inertial effects: $\rey\upi = \mathcal{O}(1)$,
	\item Weak surface tension: $\mathrm{Ca} = \mathcal{O}(\eps^{-1/2})$,
	\item Weak boundary effects: $L_\mathrm{w} = \mathcal{O}(\eps^{-1/2})$, 
	\item Weak symmetry breaking: $\phi = \mathcal{O}(\eps^{1/2})$.
\end{itemize}
Since the normalized time scale of \eqref{eq:magma} is long, $T/\eps^{1/2}$, the \emph{physical} time of validity for the conduit equation is for times much less than $T/\eps^{3/2}$.


\section{Discussion and Conclusions} \label{sec:discussion}

The key result of this perturbative approach to describing the dynamics of unsteady viscous conduits is that the conduit equation \eqref{eq:magma} is shown to be a robust model of truly dissipationless/dispersive hydrodynamics in the fully nonlinear regime, akin to a superfluid.  While nonlinear, dispersive waves are prominent in the description of free fluid interfaces, for instance in shallow water \cite{korteweg_change_1895} or in deep, stratified fluids \citep{benjamin_internal_1967,ono_algebraic_1975}, one often has to restrict to the case of weak nonlinearity to derive an analytically tractable, dissipationless, approximate model.  As we have demonstrated, the conduit equation \eqref{eq:magma} is valid in the long time, large amplitude regime, and additional physical effects must be considered for dissipation to be present in the model.  Possible sources include the use of a visco-elastic, exterior fluid \cite{grimshaw_conduit_1992} or the presence of mass diffusion across the free interface.

In addition to presenting a careful derivation of the approximate model, this article provides the theoretical basis for the experimental study of nonlinear, dispersive waves in viscous fluid conduits.  To consider the validity of our modeling assumptions and the ability of the approximate model to describe wave propagation in the system of interest, we compare with the data and scalings reported from laboratory simulations of solitary waves \citep{olson_solitary_1986,helfrich_solitary_1990}.  The key experimental parameters are given in Table \ref{tab:data}.  In both sets of experiments, it is seen that the dimensionless parameter $\eps$ is indeed small relative to unity and that the Reynolds numbers satisfy the condition for weak inertial effects.  Regarding the magnitude of the capillary number, all experiments were conducted with miscible fluids, so it is also safe to neglect the effects due to surface tension, but solubility introduces the possibility of mass diffusion.  It was reported, however, that by using an intrusive fluid that is a water-diluted version of the exterior fluid, this effect is minimal.

We can also compare our discussion from \S~\ref{sec:robustness} with the point at which the authors reported deviation between theoretically predicted solitary wave speeds and the observations.  The advantage of the multiple scales approach is that the magnitude of the small slope condition is now related directly to measurable quantities of the fluids, so it can be determined when this assumption fails.  \citet{olson_solitary_1986} reported underestimation of the wave amplitude by the approximate model at an amplitude $A\approx13$, while \citet{helfrich_solitary_1990} observed a similar underestimation when the amplitude of conduit waves exceeded $A\approx 10$.  Checking the small slope condition $\eps^{1/2} R \ll 1$ in both cases (direct evaluation gives $\eps^{1/2} R\approx$ 4.5 and 2, respectively) reveals that indeed the perturbed conduit radius exceeded the critical threshold and inertial effects would need to be incorporated to model the conduit dynamics accurately for larger amplitude.  Hence our criteria accurately predict the point of breakdown of the conduit equation as an approximate description of the interface.

With the limits of validity of the theoretical model established, viscous fluid conduits provide an optimal setting for the precise, quantitative, experimental study of DSWs.  Dispersively regularized shock waves have attracted a great deal of interest in recent years due to their observation in a range of physical systems, to include ultracold, dilute gas \citep{dutton_observation_2001,hoefer_dispersive_2006}, ion-acoustic plasma \citep{taylor_observation_1970}, nonlinear optics \citep{wan_dispersive_2007,conti_observation_2009}, and shallow water \citep{chanson_tidal_2010}, but careful comparisons of theory and data are lacking.  One difficulty is the long length and time scales required for the study of DSWs.  These slowly modulated wavetrains are characterized by the presence of two scales.  One is the $\mathcal{O}(1)$ scale of individual oscillations and the other is a long, slow scale of wave modulations $\mathcal{O}(1/\tilde{\eps})$, $\tilde{\eps} \ll 1$ (generally, $\tilde{\eps}$ is different from $\eps$).  However, images from previous experiments demonstrate that the experimental study of DSWs is accessible in viscous fluid conduits, see, e.g., Figure 18 in the review article \cite{ryan_magma_1990}.  In this setting, a DSW is created by a step-like increase in the injection rate.  This results in a larger trailing, vertically uniform conduit connected to a smaller, leading vertically uniform conduit, connected by a region of rapidly oscillating conduit waves.  By use of an automated syringe pump, high resolution imaging and accurate measurement of the fluid densities and viscosities, precise quantitative experiments are possible.  With the long-time validity of the conduit equation \eqref{eq:magma}, measurements of characteristic DSW features -- leading and trailing edge speeds and leading edge amplitude -- can be compared with the analytical results of asymptotic modulation theory \citep{whitham_non-linear_1965,el_resolution_2005}, as developed for the conduit equation \eqref{eq:magma} by the present authors in \citep{lowman_dispersive_2013}.  

The conduit equation is asymptotically equivalent to KdV in the small amplitude, long wavelength regime \citep{whitehead_korteweg-deVries_1986}, but novel DSW regimes, e.g. backflow and DSW implosion, have been observed in large amplitude numerical simulations of \eqref{eq:magma}.  This work shows that these fully nonlinear, dispersive hydrodynamic features of the reduced model are experimentally accessible in viscous fluid conduits.

\begin{table}
	\begin{center}
	\begin{tabular}{| c | c | c |}
	\hline
	 \ & Ref.~\citep{olson_solitary_1986} & Ref.~\citep{helfrich_solitary_1990} \\ \hline 
	\quad $\rho^{(e)} \ (\mathrm{g/cm^3}) \quad $ & 1.395 & 1.424  \\ \hline 
	$\rho^{(i)}  \ (\mathrm{g/cm^3})$ & 1.075 & 1.257  \\ \hline
	$\mu^{(e)} \ (\mathrm{P})$ & 110 & 45.0 \\ \hline
	$\mu^{(i)} \ (\mathrm{P}) $& 0.074 & 0.40  \\ \hline
	$L \ (\mathrm{cm})$ & 0.012 & 0.021 \\ \hline
	$U \ (\mathrm{cm/s})$ & 0.58 & $0.187 $ \\ \hline
	$T \ (\mathrm{s})$ & 0.02 & 0.11 \\ \hline
	$\eps $ & \quad $6.7 \times 10^{-4}$ \quad & \quad $8.9 \times 10^{-3}$ \\ \hline
	$Re\upi$ & 3.8 &  0.13 \\ \hline
	$Re\upe$ & \quad $3.3 \times 10^{-3}$ \quad & \quad $1.3 \times 10^{-3}$ \quad \\ \hline 
	\end{tabular}
	\end{center}
	\caption{{Experimental parameters used to study solitary waves on viscous fluid conduits in previous literature.  The parameters listed for \citet{olson_solitary_1986} are reported for conduit fluid B only because the authors reported issues due to mass diffusion with conduit fluid A and for a single background flux value for which good agreement between theory and experiment was reported.  All quantities are in cgs units with the same number of significant digits as in the original papers.}}
	\label{tab:data}
\end{table}



\emph{We thank Edward Shaughnessy for many fruitful discussions.  This work was supported by NSF Grant Nos. DGE--1252376 and DMS--1008973.}


\bibliographystyle{apsrev4-1}

\end{document}